\documentclass[final]{IEEEtran}

\usepackage{graphicx}
\usepackage{amsmath}
\usepackage{amssymb}
\usepackage{bbm}
\usepackage{subfigure}
\usepackage{bm}
\usepackage{bbold}
\usepackage{booktabs}
\usepackage{cite}
\allowdisplaybreaks[4]
\usepackage{multirow}
\usepackage{amsthm}
\usepackage{algorithm}
\usepackage{algorithmic}
\usepackage{stmaryrd}

\begin{document}
\title{GPS-Independent Localization Techniques for Disaster Rescue}

\author{Yingquan Li,~\IEEEmembership{Graduate Student Member,~IEEE}, Bodhibrata Mukhopadhyay,~\IEEEmembership{Member,~IEEE}, and Mohamed-Slim Alouini,~\IEEEmembership{Fellow,~IEEE}
\thanks{Y. Li and M. S. Alouini are with King Abdullah University of Science and Technology (KAUST), Saudi Arabia. B. Mukhopadhyay is with Indian Institute of Technology (IIT) Roorkee, India (email: yingquan.li@kaust.edu.sa, bodhibrata@ece.iitr.ac.in, slim.alouini@kaust.edu.sa).}}

\maketitle

\begin{abstract}
In this article, we present the limitations of traditional localization techniques, such as those using Global Positioning Systems (GPS) and life detectors, in localizing victims during disaster rescue efforts. These techniques usually fall short in accuracy, coverage, and robustness to environmental interference. We then discuss the necessary requirements for developing GPS-independent localization techniques in disaster scenarios. Practical techniques should be passive, with straightforward hardware, low computational demands, low power, and high accuracy, while incorporating unknown environmental information. We review various implementation strategies for these techniques, categorized by measurements (time, angle, and signal strength) and operation manners (non-cooperative and cooperative). Case studies demonstrate trade-offs between localization accuracy and complexity, emphasizing the importance of choosing appropriate localization techniques based on resources and rescue needs for efficient disaster response.
\end{abstract}

\IEEEpeerreviewmaketitle
\section{Introduction}\label{sec:intro}
Environmental disasters, encompassing earthquakes, typhoons, floods, and wildfires, result in substantial loss of life and economic damage annually. For example, the Great East Japan Earthquake in 2011, with a magnitude exceeding 9, triggered a catastrophic tsunami with waves surpassing 40 meters in height~\cite{Intro_Fukushima}. This catastrophe caused more than 15,000 fatalities and led to a severe nuclear incident at the Fukushima Daiichi power plant, which remains the only level 7 event on the International Nuclear Event Scale (INES) besides the Chornobyl disaster~\cite{Intro_Fukushima_2}. Over a decade later, the local infrastructure and ecological environment remain unrecovered. Despite the severe consequences of environmental disasters, fully mitigating their impact remains an ongoing challenge. Table~\ref{Tb:Brief} provides an overview of these events, detailing their annual occurrence and the associated human and economic losses.
\begin{table}[ht]
\caption{A brief introduction of the hazards associated with typical environmental disasters.}
\label{Tb:Brief}
\centering
\resizebox{\linewidth}{!}{\begin{tabular}{|c|c|c|c|}
\hline
\textbf{\begin{tabular}[c]{@{}c@{}}Environmental \\ disasters\end{tabular}} & \textbf{\begin{tabular}[c]{@{}c@{}}Annual \\ frequency\end{tabular}} & \textbf{\begin{tabular}[c]{@{}c@{}}Victims \\ per year\end{tabular}} & \textbf{\begin{tabular}[c]{@{}c@{}}Economic damage\\  per year\end{tabular}} \\ \hline
Earthquakes                                                                 & \textgreater 20000                             & \textgreater 10000                                             & \multirow{4}{*}{\textgreater \$1 billion}                           \\ \cline{1-3}
Typhoons                                                                    & \textgreater 1000                      & \multirow{3}{*}{\textgreater 1 million}                              &                                                                              \\ \cline{1-2}
Floods                                                                      & \textgreater 50000                                                     &                                                                      &                                                                              \\ \cline{1-2}
Wildfires                                                                   & \textgreater 1000                                                    &                                                                      &                                                                              \\ \hline
\end{tabular}}
\end{table}
%

In the aftermath of a disaster, prompt and effective rescue operations are critical for minimizing casualties and preventing further harm. 
The most significant initial step in disaster rescue is localization, which represents determining the location of victims. Current localization systems typically rely on GPS technology, which requires victims to receive satellite signals for computing location estimates~\cite{GPSPrinciple}. However, GPS localization suffers from instability and limited accuracy, especially in complex environments.

Alternatively, life detectors are employed to localize victims by identifying signs of life, such as movement, sound, or breathing, using technologies like acoustic sensors or ground-penetrating radar~\cite{LD}. While effective in some scenarios, life detectors have a notable drawback: their limited coverage area necessitates extensive manual searching, which proves impractical in large-scale disaster regions.

Given the limitations of existing localization techniques, researchers are investigating GPS-independent wireless localization techniques to enhance disaster rescue efforts. These techniques aim to provide more reliable and efficient localization services through communication between victims and rescuers, thereby overcoming the drawbacks of traditional GPS localization. In this article, we explore the methodologies underlying GPS-independent localization techniques and introduce implementation strategies that facilitate their development.

The paper is organized as follows. Section~\ref{sec:drawbacks} outlines the limitations of current localization techniques used in disaster rescue. Section~\ref{sec:requirements} discusses the requirements that localization techniques must fulfill to be effective in disaster scenarios. Section~\ref{sec:manners} introduces the implementation strategies of GPS-independent localization techniques. Section~\ref{sec:simulations} provides comparative simulations of various localization techniques applied to disaster rescue scenarios based on these strategies. Finally, we conclude this article in Section~\ref{sec:conclusion}.

\section{Limitations of current localization techniques in disaster rescue}\label{sec:drawbacks}
Since determining the location of victims is significant for facilitating further rescue operations, various localization techniques have been deployed, which can be categorized into GPS-based localization techniques and life detectors using sensing technologies. In this section, we present their respective limitations, providing a foundation for the subsequent discussion on the requirements that localization techniques suitable for disaster rescue should meet.
\subsection{GPS-based Localization Techniques}
Since the deployment of the 24-satellite system in 1993~\cite{GPSHistory}, GPS technology has been widely utilized for navigation, target tracking, and mapping. Its widespread applicability and user-friendly operation make it a common choice for disaster rescue. However, GPS localization was not specifically designed for the unique challenges of disaster rescue, and the harsh post-disaster environment further compromises its performance. The limitations of GPS-based localization techniques in disaster rescue can be summarized as follows:
\subsubsection{Active Localization} GPS localization is inherently an active process. GPS-based localization techniques necessitate that users receive signals from GPS satellites and subsequently compute their location actively. To share their location with others, users must broadcast their location information. In disaster scenarios, victims usually cannot perform this broadcasting task easily due to injury, incapacitation, or lack of equipment. Consequently, rescue personnel cannot immediately ascertain the victim's location using GPS technology alone, rendering GPS localization suboptimal for disaster rescue. This dependency on the victim's ability to actively communicate their location significantly hampers the effectiveness of GPS-based localization techniques in disaster rescue.
\subsubsection{Limited Accuracy} GPS localization lacks the necessary accuracy for effective disaster rescue. Standard consumer GPS devices typically provide localization accuracy in the range of 3 to 10 m, which can degrade further in complex post-disaster environments. This inaccuracy presents a significant challenge to rescue operations, requiring substantial resources and labor to find the victims. Inaccurate GPS Localization also leads to prolonged search times, increasing risks to victim safety, and delays in overall rescue efforts.
%
%
\subsubsection{Reliance on Satellite Signals} GPS localization systems depend heavily on stable GPS satellite signals, which poses a significant challenge in disaster rescue scenarios. Environmental disasters often lead to secondary hazards such as harsh atmospheric conditions, building shadowing, and electromagnetic interference, all of which can severely disrupt GPS signals. For instance, the 2022 volcanic eruption in Tonga produced substantial smoke and dust and damaged submarine cable systems, leading to a month-long disruption in satellite signal reception~\cite{Tonga}. In such conditions, GPS localization becomes unreliable, rendering it unsuitable for disaster rescue operations where consistent and dependable positioning information is important.

In summary, given these limitations, GPS-based localization techniques are inadequate for disaster rescue scenarios. Their active nature, reliance on stable signals, and limited accuracy present significant barriers to effective localization and rescue. These shortcomings underscore the need for alternative localization techniques that operate without GPS and better address the demands of disaster rescue.
\subsection{Life Detectors}
Another method used by rescuers to localize victims is life detectors, which identify signs of life such as movement, sound, or breathing through technologies like acoustic sensors or ground-penetrating radar. While these detectors play a significant role in rescue operations, there are three notable limitations:
\subsubsection{Lack of Accuracy} Rescuers struggle to find the exact location of victims by using life detectors. Generally, life detectors can only provide a rough area of life signs, necessitating extensive manual searches to localize victims precisely. This lack of accuracy can lead to delayed rescues, which can be fatal due to missing the golden 72-hour period post-disaster. Therefore, the limited accuracy of life detectors hinders the effectiveness of rescue efforts for survivors, making them less than satisfactory in disaster rescue.
\subsubsection{Limited Coverage} The coverage of life detectors is limited. These devices are usually effective within a few meters, requiring close proximity to the victim for detection. In large-scale disaster scenarios, such as extensive building collapses or widespread flooding, this limited coverage leads to many areas remaining unexplored, increasing the risk of missing victims. However, conducting a comprehensive survey of the disaster region demands substantial financial, equipment, and human resources, often rendering it impractical.
\subsubsection{Susceptibility to Environmental Interference} Life detectors are sensitive to environmental interference. Factors such as turbulence flow, rubble, and atmospheric electromagnetic interference can severely affect the performance of life detectors. For example, in collapsed buildings, the presence of metal materials can reduce the propagation distance of radar signals, weakening the sensors' ability to detect signs of life accurately. Additionally, these interferences can lead to false alarms and missed detections, reducing the ability of life detectors to distinguish between living and non-living targets, or between humans and animals. Therefore, environmental interference significantly undermines the reliability of life detectors, consequently affecting their effectiveness in disaster rescue efforts.

In summary, although life detectors are an important tool in disaster rescue, their lack of accuracy, limited coverage, and susceptibility to environmental interference present significant challenges. These limitations highlight the need for more reliable localization techniques to enhance the efficiency of disaster rescue operations and save more lives.

\section{Key Requirements for Localization Techniques in Disaster Rescue}\label{sec:requirements}

In Section~\ref{sec:drawbacks}, we presented the limitations of conventional localization techniques in disaster rescue. Unlike urban environments with fully established communication infrastructure or rural areas with large areas of empty space, disaster rescue scenarios impose unique requirements on localization techniques. These techniques must balance computational complexity, localization accuracy, and robustness in complex environments, while providing simultaneous services to multiple users across a wide area. Additionally, considerations of hardware architecture, such as power consumption, footprint, and operability, are significant for use in disaster rescue. In this section, we discuss the key requirements for effective localization techniques in disaster rescue, emphasizing practicality and enhancing rescue efficiency.
\begin{figure}
    \centering
    \includegraphics[width=\linewidth]{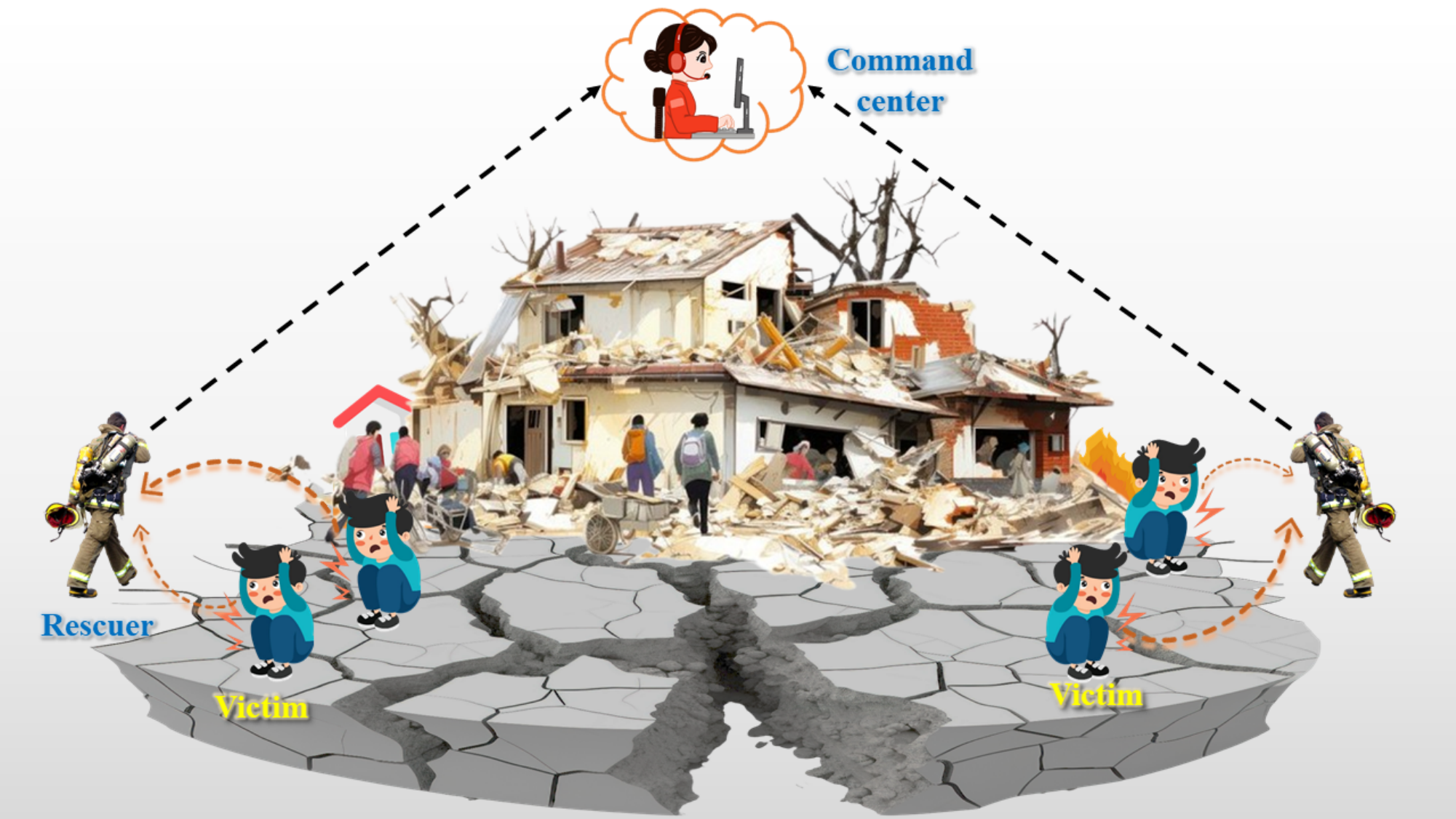}
    \caption{The procedure of passive localization in disaster rescue.}
    \label{Fig:procedure}
\end{figure}
\subsection{Passive Localization}

The localization techniques for disaster rescue should be GPS-independent, focusing on the victim's relative position to the rescuers rather than their absolute geographic coordinates on Earth. As shown in Fig.~\ref{Fig:procedure}, the victim transmits a simple signal, which is received by rescuers and forwarded to the command center. The command center collects these signals and uses localization algorithms to determine the victim's relative position to the rescuers, guiding them accordingly. This approach eliminates the need for the victim to perform complex computations or broadcast their location, as the location of the victim is not actively known by the victim himself, but passively determined by the command center. By relying on relative positioning, this approach avoids dependence on GPS signals, addressing the issues of active localization and enhancing the practicality of localization techniques in disaster rescue scenarios.
\subsection{Simple Hardware Architecture}

GPS-independent localization techniques should incorporate a simple hardware architecture. Given the unpredictable nature of environmental disasters, victims are unlikely to carry heavy equipment for potential scenarios that may never occur. Thus, GPS-independent localization techniques should be compact and lightweight, allowing them to be ideally embedded into everyday electronic devices such as mobile phones, and smartwatches. Moreover, from an energy perspective, energy constraints further necessitate a simplified hardware design. Complex hardware architectures lead to high power consumption, which is challenging in post-disaster situations where power supplies are often disrupted during disasters. High power consumption compromises long-term operation. In the aftermath of a disaster, the response time for disaster rescue varies. If the localization system consumes excessive power, it may cease functioning before rescuers arrive, thereby delaying the localization of victims and consequently increasing the risk to their lives. In addition, the necessity for complex hardware is minor since the command center manages the computational burden. On the victim's side, the localization system only needs to transmit simple signals, mitigating the need for sophisticated modems, encoders, or other digital signal processor (DSP) hardware. This makes it feasible for GPS-independent localization systems to maintain a simple, compact, and energy-efficient hardware design.
\subsection{Low Computational Complexity}

In GPS-independent localization techniques, the command center relies on optimization algorithms to determine victims' locations. It is crucial to ensure low computational complexity in these algorithms due to the time-sensitive nature of disaster rescue operations. High computational complexity can cause delays, undermining the efficiency of rescue efforts, especially within the golden 72-hour period following a disaster. Furthermore, disaster scenarios often involve limited computational resources due to infrastructure damage. Thus, low-complexity algorithms not only accelerate the localization process but also enhance the overall feasibility of the rescue operation, improving the effectiveness of GPS-independent localization techniques in disaster rescue.
\begin{figure*}
    \centering
    \includegraphics[width=\linewidth]{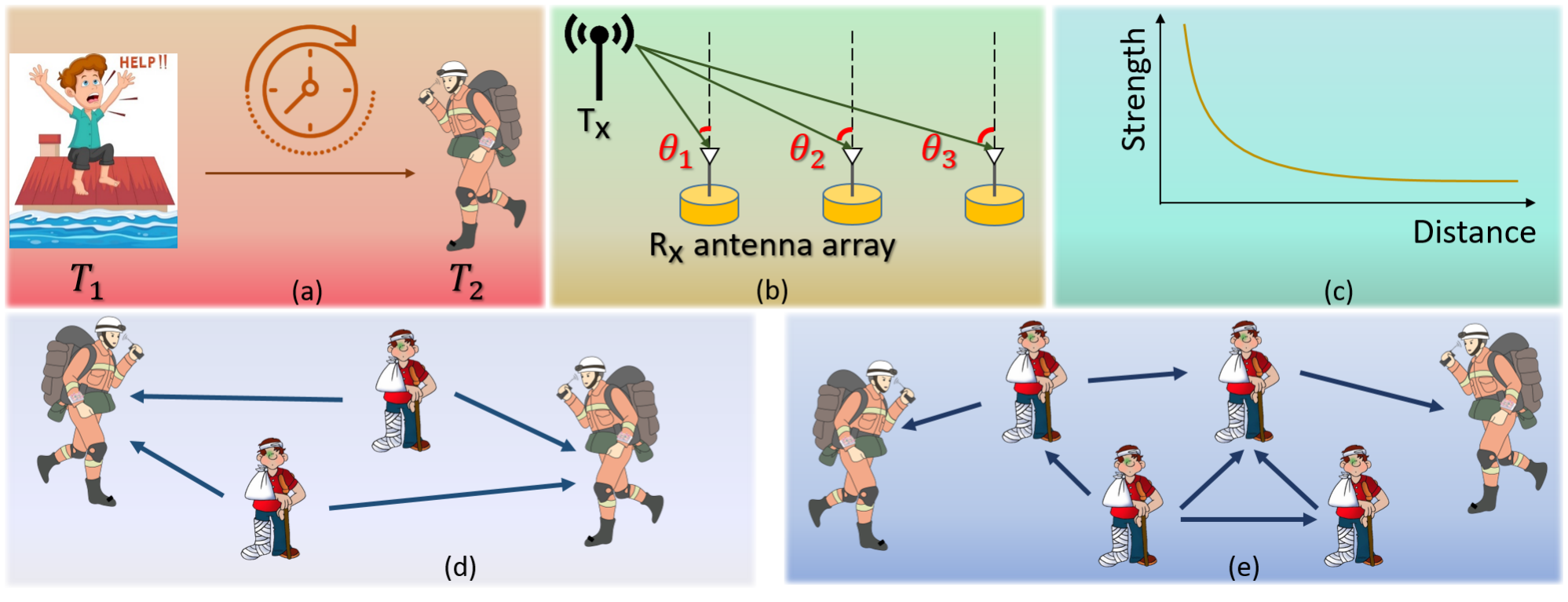}
    \caption{Implementation strategies for localization techniques: a) time-based, b) angle-based, c) strength-based, d) non-cooperative, and e) cooperative techniques.}
    \label{Fig:IS}
\end{figure*}
\subsection{High Accuracy}

In disaster rescue scenarios where every second is critical, high localization accuracy is essential for saving lives, making it a fundamental requirement for GPS-independent localization techniques. For GPS-independent localization techniques, achieving high accuracy is significant to effectively assist rescue personnel in locating and reaching victims in disaster regions. For disaster rescue, even minor localization errors can lead to fatal delays, particularly in complex environments like urban areas after earthquakes. High localization accuracy enables rescue personnel to focus their efforts effectively, minimizing wasted time and resources. This precision not only facilitates the immediate rescue of victims but also optimizes resource allocation for ongoing operations. 
\subsection{Robustness with Incomplete Information}

GPS-independent localization techniques rely on communication between victims and rescuers. Therefore, their robustness to incomplete environmental information is essential due to the unpredictable nature of disaster situations. After an environmental disaster, critical wireless channel information, such as channel capacity and path loss exponent (PLE), usually remains unknown. In addition, the impact of an environmental disaster on each victim can vary, affecting the parameters of the signal transmitter at the victim's end, including transmit power, antenna orientation, and battery level. Consequently, GPS-independent localization techniques must maintain high performance despite the lack of complete environmental information. This robustness is significant for disaster rescue, where reliable operation without comprehensive information ensures continuous and effective victim positioning and rescue.

\section{Implementation Strategies for GPS-Independent Localization Techniques}\label{sec:manners}

In this section, we introduce the implementation strategies for GPS-independent localization techniques. Fig.~\ref{Fig:IS} presents the implementation strategies reviewed in this section. Firstly, these techniques are categorized based on the type of measurements they use: time of arrival (ToA)~\cite{Intro_TOA}, angle of arrival (AoA)~\cite{Intro_AOA}, and received signal strength (RSS)~\cite{FCUP}. Some techniques also integrate these measurements, resulting in hybrid methods. Additionally, GPS-independent localization techniques can be classified by their operational manners into non-cooperative~\cite{Nonc} and cooperative approaches~\cite{Coo}. The following discussion presents an overview of the advantages and disadvantages of each type of localization technique, considering both measurements and operational manners.
\subsection{Measurements}

Measurements in GPS-independent localization techniques encompass the signal propagation characteristics obtained from the wireless links between victims and rescuers, including ToA, AoA, and RSS. As shown in Fig.~\ref{Fig:IS} a), b), and c), these measurements form the methodology for different localization techniques: ToA-based techniques estimate distances by measuring the time it takes for a signal to travel from the victim to the rescuer; AoA-based techniques determine the victim's location by analyzing the angle at which the signal arrives at the rescuer's receiver; and RSS-based techniques compute distance by using the strength of the received signal. Each technique uses a single type of measurement or a combination of measurements to provide location estimates. The subsequent subsections will introduce these measurements in detail, discussing their practicality in disaster rescue.
\subsubsection{ToA-based Techniques}

TOA-based localization techniques determine the distance between the victim and the rescuer by measuring the time it takes for a signal to travel between them~\cite{Intro_TOA}. The process involves the transmitter sending a signal with a timestamp, which is extracted by the receiver. By comparing this timestamp with the current time, the propagation delay of the signal can be obtained. Knowing the signal’s propagation speed, which is less influenced by environmental factors compared to RSS or AoA measurements, allows for accurate distance estimation. This high precision is significant in disaster scenarios, enabling the swift rescue of the victims. However, TOA-based techniques require precise synchronization between the victim and rescuer. A tiny error in timing can lead to significant location estimate error due to the high speed of electromagnetic waves. Furthermore, TOA-based techniques struggle in environments with multipath effects, where signals suffer additional delay paths. Consequently, these techniques possess sophisticated hardware and high computational complexity, which is challenging to deploy in resource-constrained disaster rescue scenarios.
Time Difference of Arrival (TDoA) is a variant of ToA. Rather than necessitating precise synchronization between the transmitter and receiver, TDoA relies on multiple synchronized rescuers to measure differences in signal arrival times. These differences are then used to estimate the location of the victim~\cite{Intro_TDOA}. This approach transfers the synchronization burden from the victim, who is in an unknown situation, to the rescuers, who are well-organized by the command center. Therefore, TDoA is practical in disaster scenarios where the requirement of synchronization for a victim is challenging. Despite these benefits, TDoA is still affected by multipath interference and has high computational complexity, which can limit its applicability in resource-limited rescue operations.
\subsubsection{AoA-based Techniques}

AoA-based techniques estimate a victim's location by measuring the angle at which a signal incident occurs in a set of receiver antennas~\cite{Intro_AOA}. By analyzing the AoA measurements relative to the known orientation of the receivers, AoA-based techniques can obtain the victim's position through triangulation. In disaster response scenarios, AoA-based techniques do not require synchronization between the victim and the rescuers. Moreover, AoA-based techniques can conduct localization with a single rescuer equipped with multiple antennas. By analyzing the variations in AoA measurements detected by these antennas, AoA-based techniques can estimate the victim’s location without needing collaboration between multiple rescuers. This capability is particularly advantageous in the early disaster response stage, where personnel is scarce, and not all rescuers are yet assembled. Nevertheless, AoA-based techniques face challenges. They typically require sophisticated equipment and specialized antenna arrays to accurately detect the signal’s incoming direction, which is expensive and cumbersome to deploy. Additionally, signal reflections common in disaster environments interfere with accurate angular measurement, reducing the performance of the location estimates. These limitations can affect the practical implementation of AoA-based techniques in dynamic and complex disaster scenarios.
\subsubsection{RSS-based Techniques}

RSS-based localization techniques leverage the relationship between distance and signal strength to estimate the distance between the victim and the rescuer~\cite{Invex}. This relationship is represented by the channel model, which varies based on the disaster area’s environment. In outdoor disaster scenarios, the log-normal channel model is commonly used. Thus, accurate channel estimation is significant before applying RSS-based techniques. However, achieving precise channel estimation across large-scale disaster regions is challenging due to the variability of wireless channels. Additionally, RSS-based localization relies on prior knowledge of the victim's signal transmit power, which can vary due to differences in battery levels or antenna orientations among victims. Therefore, it is impractical to assume that the transmit power of each victim is known. Due to the high dependence of the RSS measurement on additional environmental information, RSS-based techniques generally have poorer performance compared to ToA-based and AoA-based techniques. However, RSS-based localization benefits from its ease of implementation and minimal hardware requirements, making it cost-effective and compatible with everyday devices, such as mobile phones. Due to the lower costs and power consumption of RSS-based techniques, they are suitable for widespread deployment in remote or less-developed disaster-prone areas. Their low power consumption contributes to long-term operations in disaster scenarios, aiding victims until rescue personnel arrive and thus increasing the possibility of successful rescue.
A variant of RSS is RSS difference (RSSD) measurement, which addresses the reliance on prior knowledge of transmit power by subtracting RSS measurements received from the same victim~\cite{FCUP}. However, RSSD-based techniques are affected by additive Gaussian noise, which is amplified during subtraction. Furthermore, due to differing transmit power among victims, RSSD-based techniques can only perform individual localization sequentially rather than simultaneously locating multiple victims. Consequently, RSSD-based techniques are suitable for moderate disaster scenarios with few victims and stable wireless channels.
\subsection{Operational Manners}

The implementation of GPS-independent localization techniques can be categorized into two manners: non-cooperative and cooperative. Non-cooperative localization techniques rely solely on measurements between a single victim and rescuers, without requiring communication among victims. Each victim is localized independently, and the location estimation for one victim does not involve data from other victims. In contrast, cooperative localization techniques include not only communication between victims and rescuers but also wireless links between victims to enhance the accuracy and robustness of the localization process. Victims in a cooperative system can transmit and receive signals from each other, which are then forwarded to the command center to provide comprehensive location information for multiple victims simultaneously. The choice between non-cooperative and cooperative localization techniques in disaster rescue scenarios depends on factors such as resource availability, environmental conditions, and the specific needs of the rescue operation. The following sections provide an overview of each manner within the context of disaster rescue scenarios.
\begin{table*}[ht]
\caption{A brief overview of existing localization techniques.}
\label{Tb:Techniques}
\centering
\begin{tabular}{|c|c|c|c|c|}
\hline
\textbf{Techniques} & \textbf{Measurements} & \textbf{Manners} & \textbf{CPU runtime (in s)} & \textbf{Year} \\ \hline
ToA-Chen~\cite{Intro_TOA}            & ToA                   & Cooperative      & 94.71                       & 2020          \\ \hline
EM-POG-AMP~\cite{Intro_AOA}          & AoA                   & Cooperative      & 17.75                       & 2022          \\ \hline
P-TDoA~\cite{Intro_TDOA}              & TDoA                  & Non-cooperative  & 153.06                      & 2020          \\ \hline
RLBM~\cite{Lohrasbipeydeh_Gulliver_2021}                & RSSD                  & Non-cooperative  & 3.88                        & 2021          \\ \hline
IRDL~\cite{Invex}                & RSS                   & Cooperative      & 0.80                        & 2022          \\ \hline
FCUP~\cite{FCUP}                & RSS                   & Cooperative      & 2.68                        & 2023          \\ \hline
\end{tabular}
\end{table*}
\subsubsection{Non-Cooperative Localization Techniques}

Non-cooperative localization techniques are typically more straightforward to implement, as they do not require additional communication links between victims~\cite{Nonc}. This simplicity is advantageous in disaster scenarios where rapid deployment is essential. With reduced requirements for synchronization and data sharing, non-cooperative techniques spend lower hardware costs. Consequently, these techniques are budget-friendly, making them suitable for scenarios with limited resources. A significant drawback is that non-cooperative techniques can only localize victims individually and also require the victim to be connected to multiple rescuers, rather than simultaneously determining the positions of multiple victims. This limitation becomes critical when the number of victims is high, as non-cooperative techniques need to consume extensive time to find all victims. The extended rescue time can decrease the overall efficiency of disaster rescue efforts and lead to considerable resource consumption, reducing the cost-effectiveness of these techniques.
\subsubsection{Cooperative Localization Techniques}

As shown in Fig.~\ref{Fig:IS} e), cooperative localization techniques enable the joint processing of data from multiple victims, allowing for the simultaneous determination of their locations and thereby improving overall system performance~\cite{Coo}. This capability is particularly valuable in disaster scenarios where precise localization is crucial for effective rescue operations. In cooperative localization, victims who cannot directly communicate with enough rescuers can still relay their information indirectly through links with other victims, enabling their localization by rescuers. Despite these advantages, cooperative localization techniques have high complexity due to the need for synchronization, data sharing, and coordination among multiple victims and rescuers. This increased complexity requires more sophisticated hardware and computational loads, which can be challenging to maintain in resource-limited disaster scenarios. Furthermore, leveraging wireless links between victims introduces potential risks associated with communication failures. The damage to communication infrastructure resulting from the disaster severely affects the victims' communication capabilities. Consequently, measurements obtained via victim-victim links can become unreliable, bringing numerous amounts of abnormal data and thereby degrading the reliability of cooperative localization techniques.

\section{Case Study: Localization performance of techniques with different implementation strategies}\label{sec:simulations}

In this section, we perform comprehensive numerical simulations to provide a comparative evaluation of existing GPS-independent localization techniques, focusing on both performance and computational complexity. The considered techniques~\cite{Intro_TOA,Intro_AOA,Intro_TDOA,Lohrasbipeydeh_Gulliver_2021,Invex,FCUP} and their corresponding measurements are presented in Table~\ref{Tb:Techniques}. For iterative techniques, i.e. RLBM~\cite{Lohrasbipeydeh_Gulliver_2021} and IRDL~\cite{Invex}, the convergence is achieved when the absolute difference between objective function values in successive iterations is less than $10^{-3}$. Due to its high computational complexity associated with semidefinite programming (SDP), RLBM is limited to a maximum of 10 iterations, whereas IRDL allows up to 2000 iterations. SDP-based techniques are implemented using the SeDuMi solver within the CVX toolbox, with precision set to ``best." Hyperparameters align with those provided in corresponding references. We carry out 3000 Monte Carlo simulations for each study to evaluate the estimation accuracy of victims' locations, using the normalized root mean squared error (NRMSE) as the performance metric. The average CPU runtime per Monte Carlo simulation measures the computational complexity of each technique. The disaster region is simulated by an outdoor network of 5 victims and 10 rescuers, randomly distributed within a $100\times100~m^2$ area with fixed locations throughout the simulations. Victims' transmit powers are randomly selected from a uniform distribution between [-10 10]~dBm, and the PLE for wireless channels is set to 3.
\begin{figure}
    \centering
    \includegraphics[width=\linewidth]{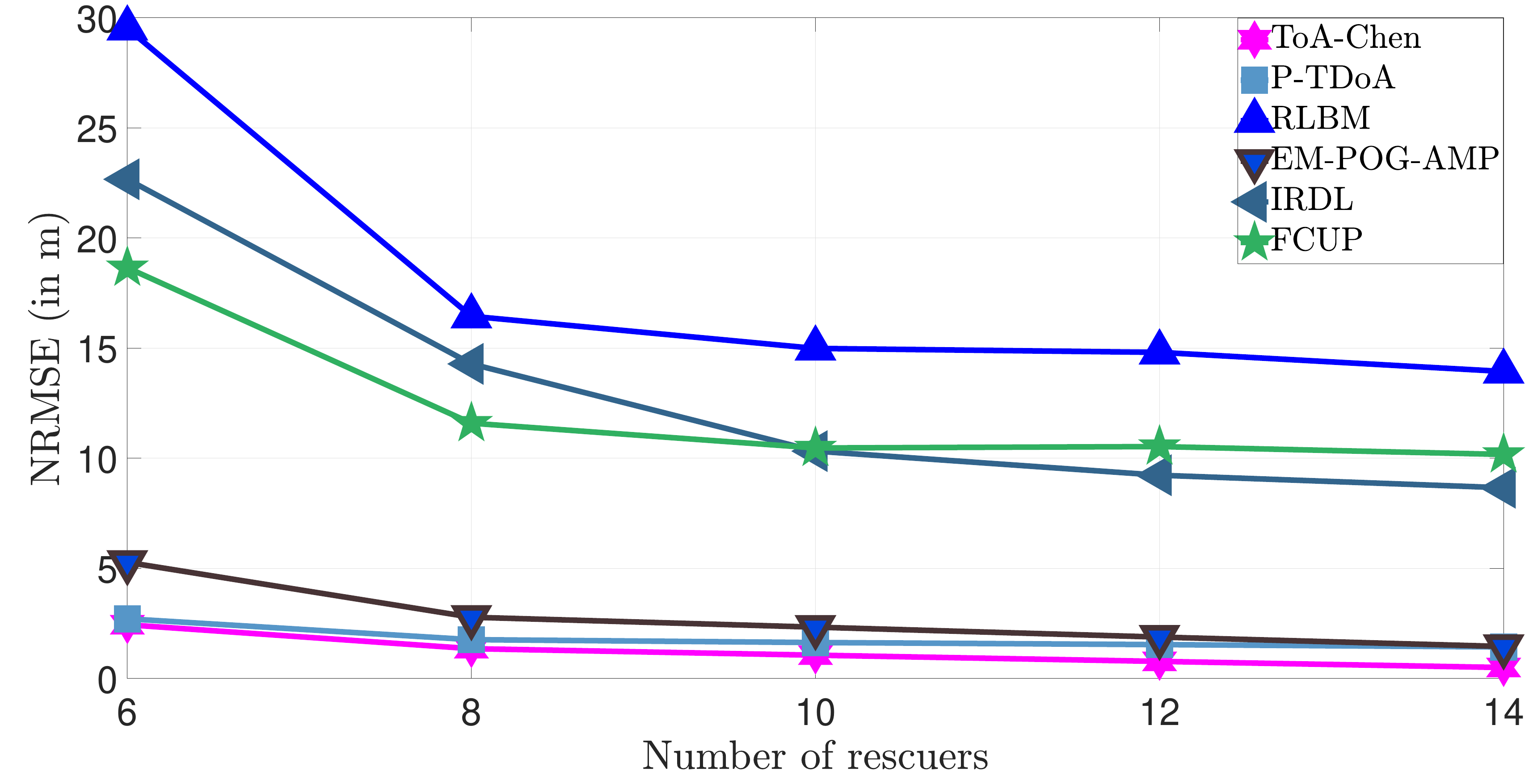}
    \caption{Performance of the localization techniques for estimating victims' locations as a function of the number of rescuers.}
    \label{Fig:Na}
\end{figure}
\begin{figure}
    \centering
    \includegraphics[width=\linewidth]{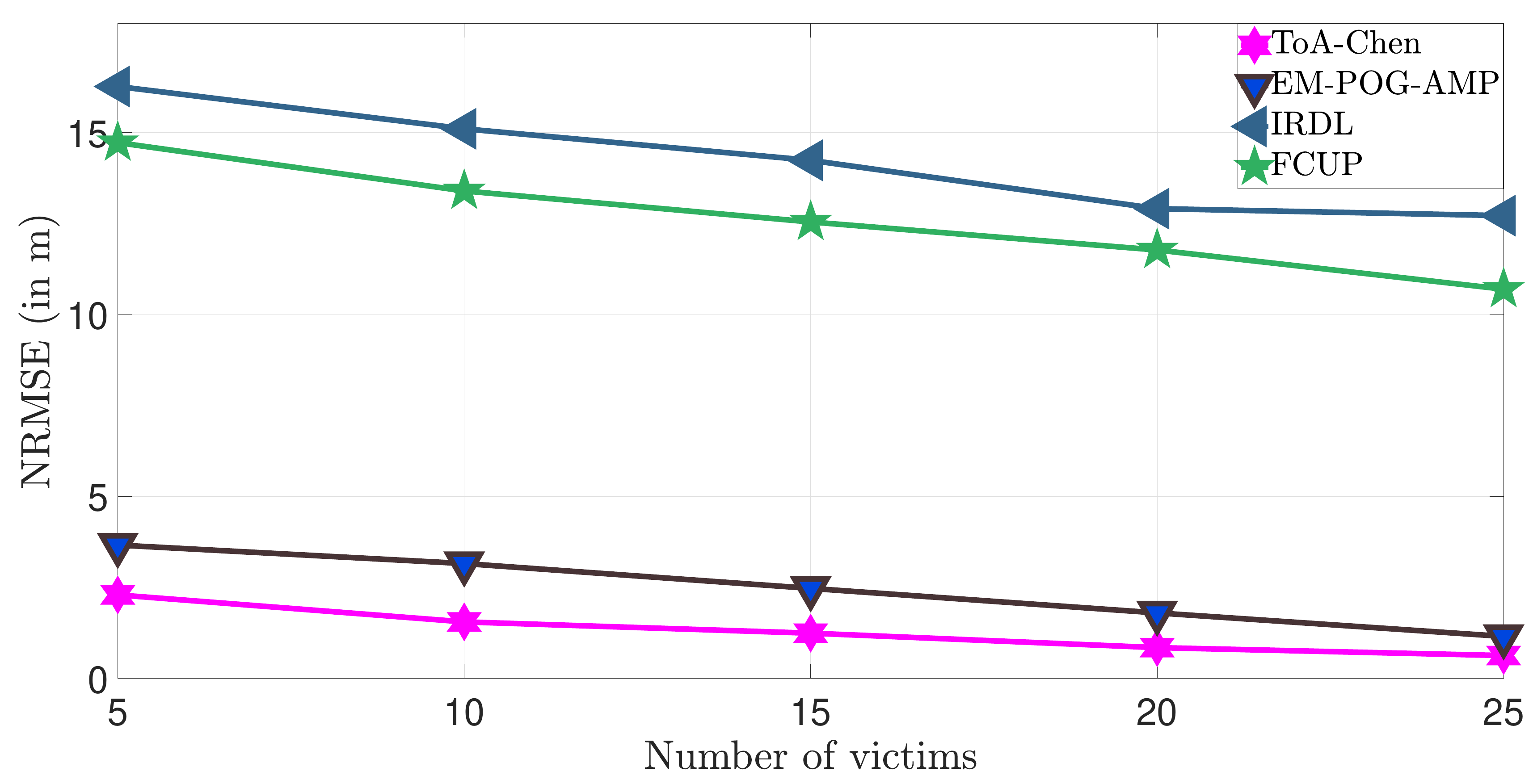}
    \caption{Performance of the localization techniques for estimating victims' locations as a function of the number of victims.}
    \label{Fig:Nt}
\end{figure}
\subsection{Effect of the Number of Rescuers}

This study evaluates the performance of various localization techniques with different measurements as the number of rescuers increases from 6 to 14. The results are depicted in Fig.~\ref{Fig:Na}, which shows the NRMSE for victim location estimation. It can be observed that ToA-Chen and P-TDoA exhibit superior performance compared to other techniques. Specifically, ToA-Chen marginally outperforms P-TDoA. This priority results from the cooperative nature of ToA-Chen, which leverages additional links between victims to enhance localization accuracy. In contrast, however, P-TDoA shifts the synchronization burden from victims to rescuers, making it more practical in real-world disaster scenarios where victims are in unpredictable conditions. EM-POG-AMP demonstrates suboptimal performance with fewer than 8 rescuers but achieves performance comparable to time-based techniques when the number of rescuers exceeds 8. Techniques using RSS measurements perform notably worse than those using time or angle measurements, exhibiting significantly higher NRMSE values. This discrepancy highlights the limitations of RSS-based techniques, which are heavily influenced by environmental factors. The performance of RLBM, which does not utilize victim-victim links, is outperformed by IRDL and FCUP. From a perspective of location estimation accuracy, techniques employing time or angle measurements generally provide better performance, achieving localization accuracy within 1~m when an adequate number of rescuers is available.
\subsection{Effect of the Number of Victims}

Fig.~\ref{Fig:Nt} illustrates the performance of the localization techniques as the number of victims increases from 5 to 25 within the disaster area. We exclude non-cooperative techniques, i.e. P-TDoA and RLBM, from this study due to their inability to leverage victim-victim links for localization. Additionally, non-cooperative techniques require at least three rescuers to communicate with each victim for effective location estimation. However, this condition becomes difficult to ensure with numerous victims in a network. Fig.~\ref{Fig:Nt} demonstrates that the performance of cooperative localization techniques improves with an increasing number of victims. This improvement is due to the additional measurements available from the extra victim-victim links. In disaster scenarios, where rapidly localizing a large number of victims is necessary, the ability of cooperative localization techniques to simultaneously estimate the locations of multiple victims is advantageous. Among these techniques, ToA-Chen stands out for its superior performance, further emphasizing the effectiveness of using time-based measurements in disaster rescue operations. 
\subsection{Complexity comparison}\label{subsec:complexity}
In Table~\ref{Tb:Techniques}, we compute the average CPU runtime of one Monte Carlo simulation as a performance metric of the complexity of all techniques in MATLAB R2021a using an Intel(R) Xeon(R) W-2245 CPU @ 3.90GHz processor with 64GB RAM. While ToA-Chen provides the highest localization accuracy, it also exhibits high computational complexity, surpassing techniques based on AoA or RSS measurements. The non-cooperative nature of P-TDoA necessitates sequential localization of victims, resulting in the longest processing time. Conversely, RSS-based techniques generally present lower complexity. IRDL’s minimal CPU runtime, achieved through low-complexity optimization algorithms like gradient descent, makes it particularly suitable for urgent localization tasks in disaster rescue scenarios. Overall, the results of these studies highlight the need to choose appropriate localization techniques based on the number of rescuers, the number of victims, and the computational resource constraints of the rescue operation. For scenarios demanding high accuracy and where resources are adequate, time-based or AoA-based techniques should be prioritized. However, when budget constraints are tight and immediate victim rescue is critical, RSS-based techniques should be employed initially to quickly find an approximate range of victims’ locations, facilitating immediate rescue efforts. If feasible, more accurate localization techniques can be applied subsequently to enhance rescue efforts.

\section{Conclusion}\label{sec:conclusion}

In this article, we have presented the shortcomings of conventional GPS-based localization and life detectors in disaster rescue operations. We have discussed the key requirements for developing GPS-independent localization techniques applicable to disaster rescue. We have outlined various implementation strategies, focusing on measurements and operational manners, with a discussion of their respective advantages and disadvantages. Case studies demonstrated that time-based and angle-based techniques offer superior localization accuracy, while RSS-based methods provide benefits in terms of complexity. Therefore, the selection of optimal techniques for disaster rescue should consider disaster situations, costs, and specific needs to enhance rescue efficiency.

\bibliography{IEEEabrv,main}
\bibliographystyle{IEEEtran}

\section*{biographies}
\begin{IEEEbiographynophoto}
{Yingquan Li} is a Ph.D. candidate at King Abdullah University of Science and Technology (KAUST), Thuwal, Saudi Arabia. 
\end{IEEEbiographynophoto}
\begin{IEEEbiographynophoto}
{Bodhibrata Mukhopadhyay} is an assistant professor at the IIT Roorkee.
\end{IEEEbiographynophoto}
\begin{IEEEbiographynophoto}
{Mohamed-Slim Alouini} (Fellow, IEEE) is a distinguished professor at KAUST.
\end{IEEEbiographynophoto}

\end{document}